\documentclass[a4paper]{jpconf}
\usepackage{graphicx}
\usepackage{iopams}

\usepackage{amsmath}
\usepackage{placeins}

\usepackage{hyperref}

\def\eq#1{(\ref{#1})}
\def\Eq#1{Eq.~(\ref{#1})}
\def\Fig#1{Fig.~\ref{#1}}
\def\Ref#1{Ref.~\cite{#1}}
\def\Refs#1{Refs.~\cite{#1}}

\newcommand{\be}{\begin{equation}}
\newcommand{\ee}{\end{equation}}
\newcommand{\bea}{\begin{eqnarray}}
\newcommand{\eea}{\end{eqnarray}}

\def\blr#1{\left(#1\right)}
\def\slr#1{\left[#1\right]}

\def\mc#1{\mathcal{#1}}
\newcommand{\mrm}[1]{\mathrm{#1}}
\def\Nc{N_\mrm{c}}
\def\pd{\partial}
\def\Phib{\bar{\Phi}}
\def\qbq{\mrm{\bar{q}q}}

\graphicspath{{./fig/}}

\begin{document}

\title{Aspects of isentropic trajectories in chiral effective models}

\author{Rainer Stiele$^{1,2}$, Wanda Maria Alberico$^{1,3}$, Andrea Beraudo$^1$, Renan C\^amara Pereira$^4$, Pedro Costa$^4$, Hubert Hansen$^5$ and Mario Motta$^{1,3}$}
\address{$^1$ INFN -- Sezione di Torino, Via Pietro Giuria 1, I-10125 Torino, Italy}
\address{$^2$ Univ.~Lyon, ENS de Lyon, Univ Claude Bernard Lyon 1, CNRS, F-69342 Lyon, France}
\address{$^3$ Dipartimento di Fisica, Universit\`a degli Studi di Torino, Via Pietro Giuria 1, I-10125 Torino, Italy}
\address{$^4$ CFisUC, Department of Physics, University of Coimbra, P-3004 - 516  Coimbra, Portugal}
\address{$^5$ Univ.~Lyon, Universit\'e Claude Bernard Lyon 1, CNRS/IN2P3, IP2I-Lyon, F-69622, Villeurbanne, France}

\ead{rainer.stiele@ens-lyon.fr}

\begin{abstract}
	The evolution of the fireball in heavy ion collisions is an isentropic process, meaning that it follows a trajectory of constant entropy per baryon in the phase diagram of the strong interaction.
	Responsible for the collective acceleration of the fireball is the speed of sound of the system, while fluctuations of conserved charges are encoded in quark-number susceptibilities: together, they leave their imprint in final observables.
	Here, this isentropic evolution will be analysed within chiral effective models that account for both chiral and center symmetry breaking, two central aspects of QCD. Our discussion focusses on the impact on the isentropic trajectories of the treatment of high-momentum modes, of the meson contribution to thermodynamics and of the number of quark flavours.
\end{abstract}

\FloatBarrier

\section{Introduction}

Chiral effective models allow one to study two important properties of QCD, namely the confinement of constituent quarks with large effective masses at low temperature/density and their deconfinement to light quarks in the hot/dense partonic phase. 
Generation of constituent quark masses $m_f$ can be described by dynamical breaking of chiral symmetry and deconfinement by spontaneous breaking of center symmetry. 
The interaction between constituent quarks can be described either as a point-like interaction or by the exchange of a meson.
The former leads to what is called the Nambu--Jona-Lasinio (NJL) model \cite{Nambu:1961tp,Nambu:1961fr} and the latter to the Quark-Meson (QM) model \cite{GellMann:1960np}.
Coupled to the order parameter of center symmetry breaking, the Polyakov loop $\Phi$, these models allow one to explore phenomenologically the phase diagram of strongly-interacting matter \cite{Mocsy:2003qw,Fukushima:2003fw}.
Their Lagrangian can be written as a part describing the spontaneous breaking of chiral symmetry, a kinetic contribution of the quarks that contains a minimal coupling to the gauge field $A_\mu$ and the Polyakov-loop potential $\mc{U}$
\be
	\mc{L}_\mrm{PNJL/PQM} = \mc{L}_\mrm{chiral} +\bar{q}\slr{i\gamma_{\mu}\left(D^{\mu} + \widehat{\mu_f}\,\delta^{\mu0}\right)} q
	-\mathcal{U}\left(\Phi\!\slr{A_\mu},\Phib\!\slr{A_\mu};T\right)\,.
\label{eq:LagrangianPNJLPQM}
\ee
While the first part differs in the NJL and QM models, the other two contributions are common to both \cite{Hansen:2019lnf,Pereira:2019_1}.
This structure is conserved for the grand canonical potential in mean-field approximation,
\be
	\Omega\blr{m_f,\Phi,\Phib;\,T,\mu_f} = U_\mrm{chiral}\blr{m_f}
	+ \Omega_\qbq\blr{m_f,\Phi,\Phib;\,T, \mu_f}
	+ \mc{U}\blr{\Phi,\Phib;\,T} \;.
	\label{eq:grand_canon_pot}
\ee
All thermodynamic quantities can be derived from \Eq{eq:grand_canon_pot} ensuring the Gibbs-Duhem relation,
\be
	p = -\Omega\;, \quad s = \left.\frac{\pd{p}}{\pd{T}}\right|_{\mu_f=\mrm{const}}\;, \quad n_f = \left.\frac{\pd{p}}{\pd\mu_f}\right|_{T=\mrm{const}}\; \quad \mrm{and} \quad \epsilon  = Ts - p + \sum_{N_f} \mu_f\, n_f\;.
\ee
Here, several results for the trajectories with constant $s/n_\mrm{q}$ ratio, corresponding to the isentropic evolution of the matter produced in relativistic heavy-ion collisions, will be discussed\footnote[1]{Indeed, for AGS, SPS, and RHIC, the values of $s/n_\mrm{B}$ are, respectively, 30, 45, and 300 \cite{Bluhm:2007nu}. At these values, lattice results for the isentropic (2+1)-flavour equation of state were obtained in Refs. \cite{DeTar:2010xm,Borsanyi:2012cr}.}.
An in-depth analysis together with the evaluation along these trajectories of thermodynamic quantities like the speed of sound -- responsible for the collective acceleration of the fireball -- and generalised quark-number susceptibilities -- connected to the fluctuations of conserved charges -- will be given in \Refs{Pereira:2019_1,Motta:2019}.

\section{Impact of the UV cutoff}

\begin{figure}
	\centering
	\includegraphics[width=0.496\textwidth]{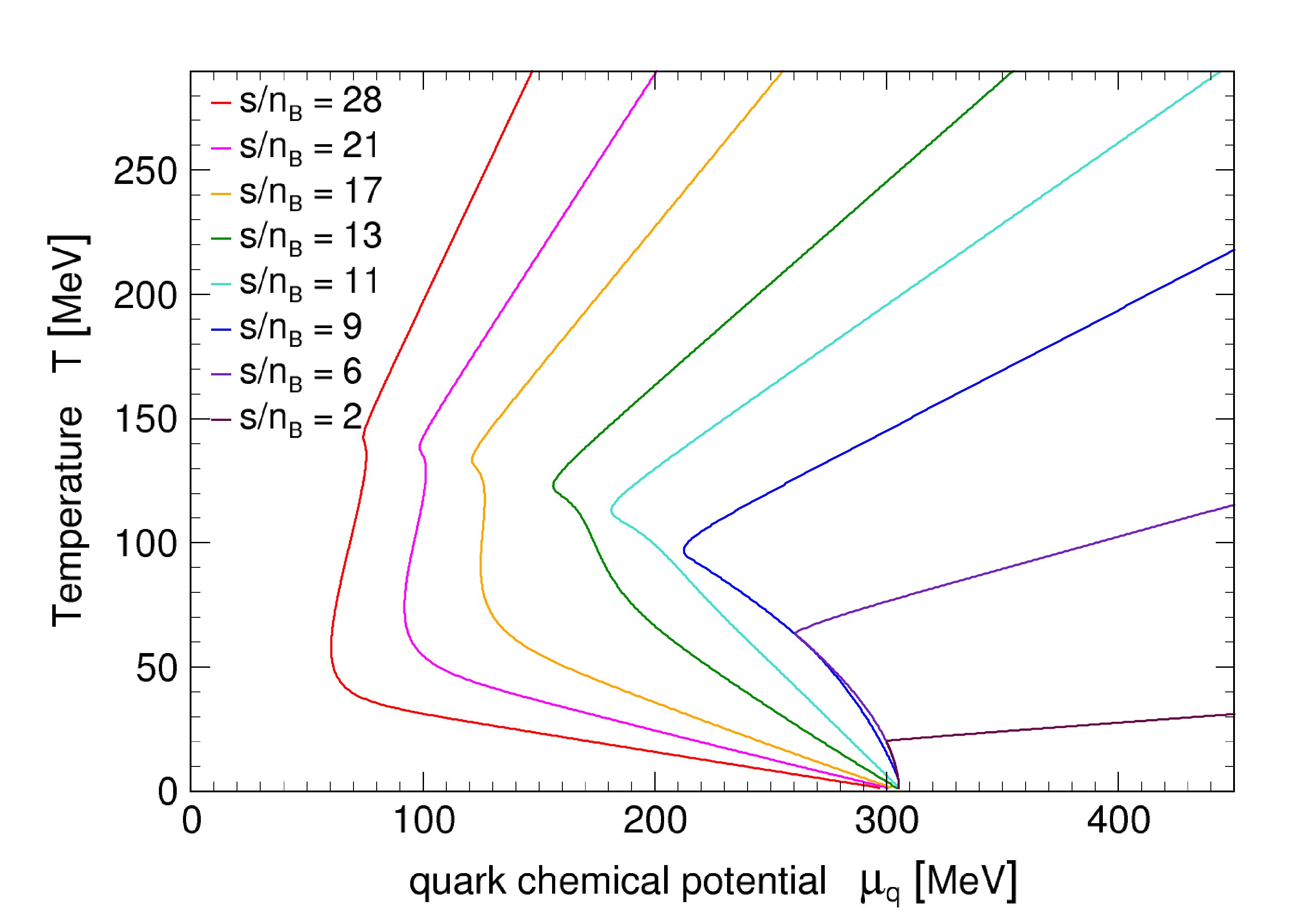}
	\hfill
	\includegraphics[width=0.496\textwidth]{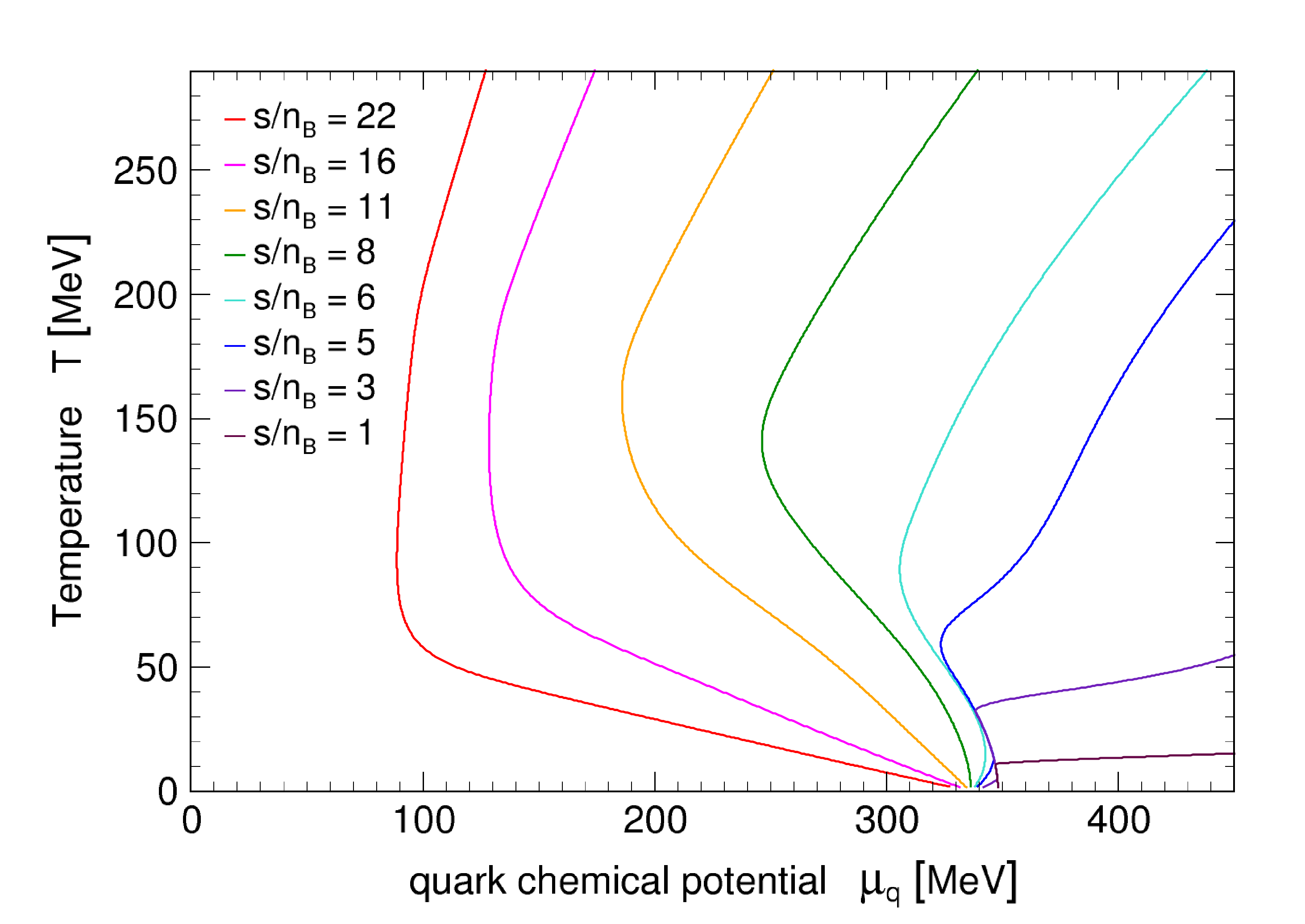}
	\caption{
	Isentropic lines in the $N_\mrm{f}\!=\!2$ Quark-Meson model (left panel) and NJL model (right panel) as in Fig.~9 of \Ref{Scavenius:2000qd}.}
	\label{fig:isentropes_QM_NJL_Nf2}
\end{figure}

Figure \ref{fig:isentropes_QM_NJL_Nf2} shows isentropic lines in the QM model (left panel) and NJL model (right panel), as presented in \Ref{Scavenius:2000qd}, see also \Ref{Kahara:2008yg}.
Isentropic trajectories look quite different in the two models: we wish to analyse such an issue more deeply.

\begin{figure}
	\begin{minipage}[b]{0.496\textwidth}
		\caption{Impact of the UV cutoff on the isentropic trajectories in the NJL model.
		The dashed curves are the ones shown in the right panel of \Fig{fig:isentropes_QM_NJL_Nf2} and are obtained applying the UV cutoff to all momentum integrals.
		The solid curves are obtained applying the UV cutoff only to the divergent integral over the vacuum fluctuations.
		In this case, the shape of the lines of constant entropy per quark is closer to those of the Quark-Meson model, shown in the left panel of \Fig{fig:isentropes_QM_NJL_Nf2}.}
		\label{fig:isentropes_NJL}
	\end{minipage}
	\hfill
	\includegraphics[width=0.496\textwidth]{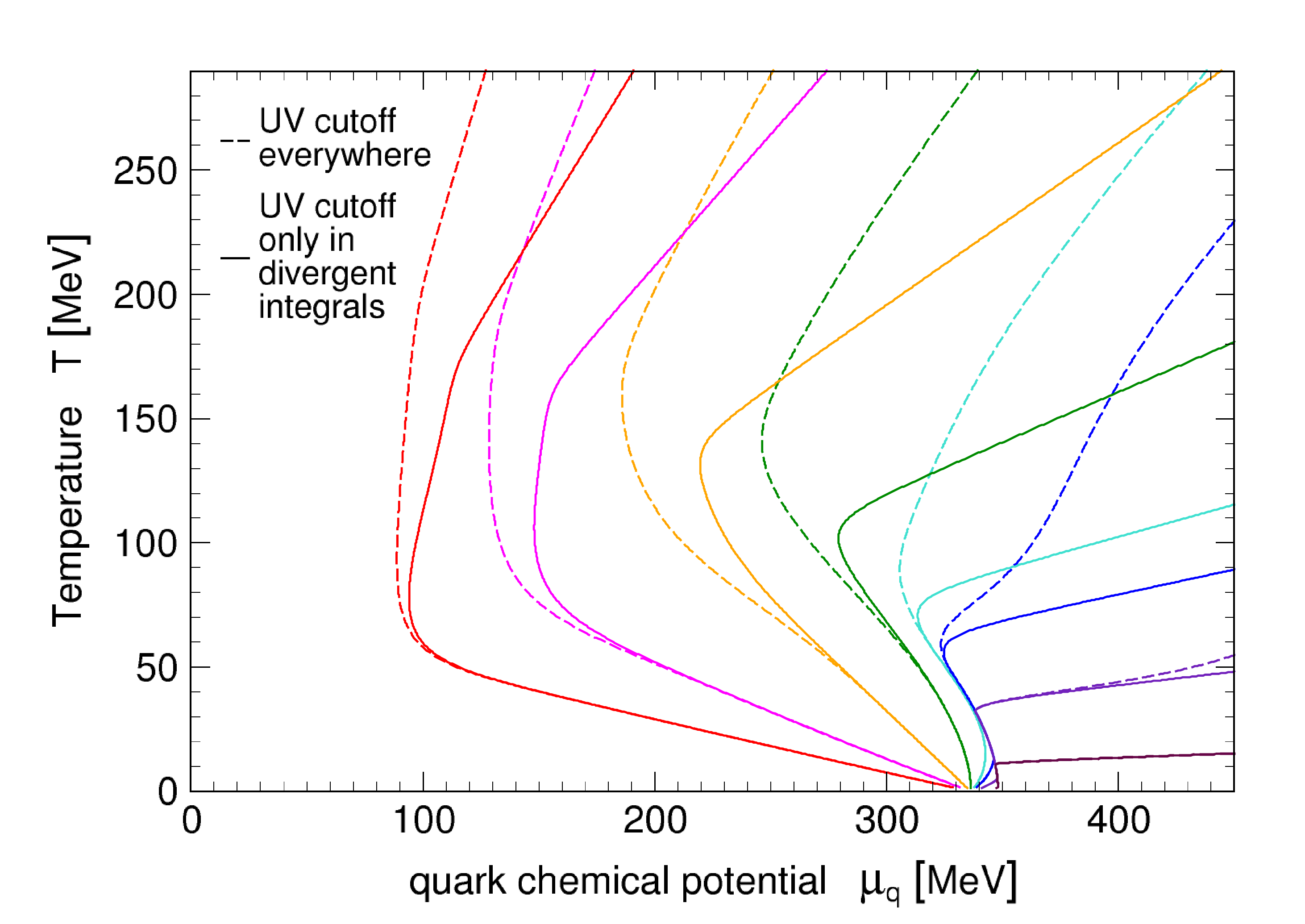}
\end{figure}

The quark contribution to the grand canonical potential \eq{eq:grand_canon_pot} can be written as the sum of two contributions. One is a vacuum-like term that does not depend \emph{explicitely} on medium properties like temperature or chemical potential,
\be
	\Omega_\qbq^\mrm{vac}\blr{m_\ell,m_\mrm{s}} = -2\Nc\sum_{f}\int\limits^\Lambda\!\frac{\mrm{d^3}k}{(2\pi)^3}\,E_f \overset{\mrm{QM}}{=} -\frac{\Nc}{8\pi^2} \sum_{f} m_f^4\, \ln\blr{\frac{m_f}{\Lambda}} \;.
	\label{eq:QuarkVac}
\ee
In the above the integral over the three-momentum is divergent and requires an ultra-violet cutoff $\Lambda$. However, the divergence can be renormalised in the QM model \cite{Skokov:2010sf}.\footnote{The formal dependence of \Eq{eq:QuarkVac} on the cutoff is cancelled in the renormalised QM model by a formal dependence of the parameters of $U_\text{chiral}^\text{QM}$ on the cutoff such that \Eq{eq:grand_canon_pot} and derived equations are cutoff independent.}
The other contribution contains the quark and anti-quark distribution functions that are defined e.g.~in \cite{Hansen:2019lnf}
\be
    \Omega_\qbq^\mrm{th}\blr{m_f,\Phi,\Phib;\,T, \mu_f}=-2T\sum_f\int\limits^{\Lambda,\infty}\frac{\mrm{d^3}k}{(2\pi)^3}\slr{z_f\!\blr{m_f,\Phi,\Phib;\,T, \mu_f}+z_{\bar{f}}\!\blr{m_f,\Phi,\Phib;\,T, \mu_f}}\;.
 \label{eq:QuarkTh}
\ee
It is therefore a convergent integral and in the QM model there is no reason to apply a UV cutoff. In the NJL model the situation is different.
One can either define the whole theory with a UV cutoff and apply the one of \eq{eq:QuarkVac} also to \eq{eq:QuarkTh} as is done for the NJL result in \Fig{fig:isentropes_QM_NJL_Nf2}.
Otherwise, in order not to lose a relevant contribution from high-momentum modes to thermodynamics, one can apply the UV cutoff only to the divergent integral \eq{eq:QuarkVac}.
Figure \ref{fig:isentropes_NJL} shows the impact of applying the cutoff $\Lambda$ to the non-divergent integrals: the shape of the lines of constant entropy per particle in the NJL model is closer to the one of the QM model when integrating up to infinity in all convergent thermal integrals.
A detailed discussion can be found in \Ref{Costa:2009ae}.

\begin{figure}
	\includegraphics[width=0.496\textwidth]{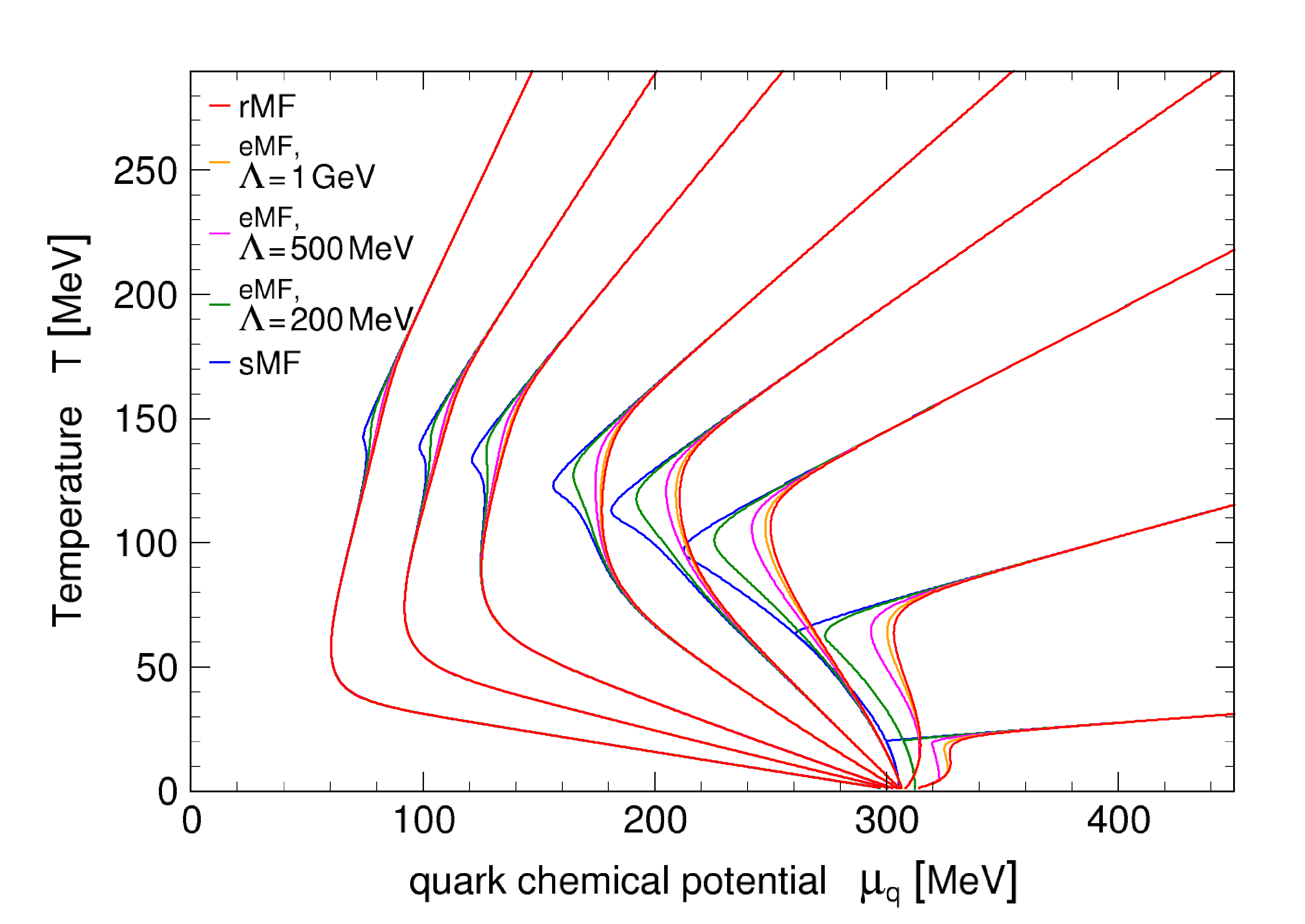}
	\hfill
	\begin{minipage}[b]{0.496\textwidth}
		\caption{Impact of the cutoff scale in \eq{eq:QuarkVac} on the isentropic trajectories in a Quark-Meson model calculation. The blue lines are the ones shown in the left panel of \Fig{fig:isentropes_QM_NJL_Nf2}.}
		\label{fig:isentropes_cutoff}
	\end{minipage}
\end{figure}

Furthermore, the cutoff in \Eq{eq:QuarkVac} should be chosen large enough to have little quantitative impact on the results.
The QM model allows one to study how the results converge when increasing the UV cutoff scale from zero (considered as the standard mean-field analysis in the QM model, applied in \Ref{Scavenius:2000qd} to obtain the results in the left panel of \Fig{fig:isentropes_QM_NJL_Nf2}) towards the renormalised mean-field model, see \Fig{fig:isentropes_cutoff}.

\section{Impact of contribution of mesons to thermodynamics \label{sec:MesonThd}}

In the common mean-field approximation of (P)QM/(P)NJL models only thermal fluctuations of quarks contribute to the thermodynamic potential, while any thermal contribution from the mesons is neglected.\\
To achieve a better description of thermodynamics in the phase where mesons are the dominant degrees of freedom, the thermodynamics can be augmented by the contribution of a gas of thermal mesons \cite{Haas:2013qwp,Herbst:2013ufa}. The contribution to the pressure of each meson species $\phi_i$ is
\be
	p_{\phi_i} = \frac{1}{\blr{2\pi}^3} \int_0^\infty \mrm{d^3}k\,\frac{k^2}{3 E_{\phi_i}}\,\frac{1}{e^{(E_{\phi_i}-\mu_{\phi_i})/T}-1} \qquad \mrm{with} \qquad E_{\phi_i}=\sqrt{k^2+m_{\phi_i}^2} \;.\qquad
	\label{eq:MesonPressure}
\ee
The total contribution of the mesons to the pressure $p_\phi$ is accordingly $p_\phi=\sum_i p_{\phi_i}$ and overall $p = -\Omega + p_\phi$.
The meson masses that enter into the dispersion relations of the mesons in \Eq{eq:MesonPressure} are medium dependent.
This medium dependence of the meson masses makes the pressure of the mesons $p_\phi$ strictly speaking a field($\psi_i$)-dependent correction to the thermodynamical potential that contributes as well to the equations of motion, $\pd\Omega/\pd\psi_i-\pd{p_\phi}/\pd\psi_i$ and the meson masses themselves.
In a lowest order approximation this correction could be neglected, assuming that the dynamics of the system is governed by the grand canonical potential $\Omega$ alone which determines the meson masses as well.
An uncoupled meson gas with these meson masses is then added to thermodynamics via $p = -\Omega + p_\phi$. 
All thermodynamic quantities are derived from this equation 
ensuring the Gibbs-Duhem relation.
All contributions stemming from $\pd{p_\phi}/\pd\psi_i$ are neglected there consistently as well.

\begin{figure}
	\includegraphics[width=0.496\textwidth]{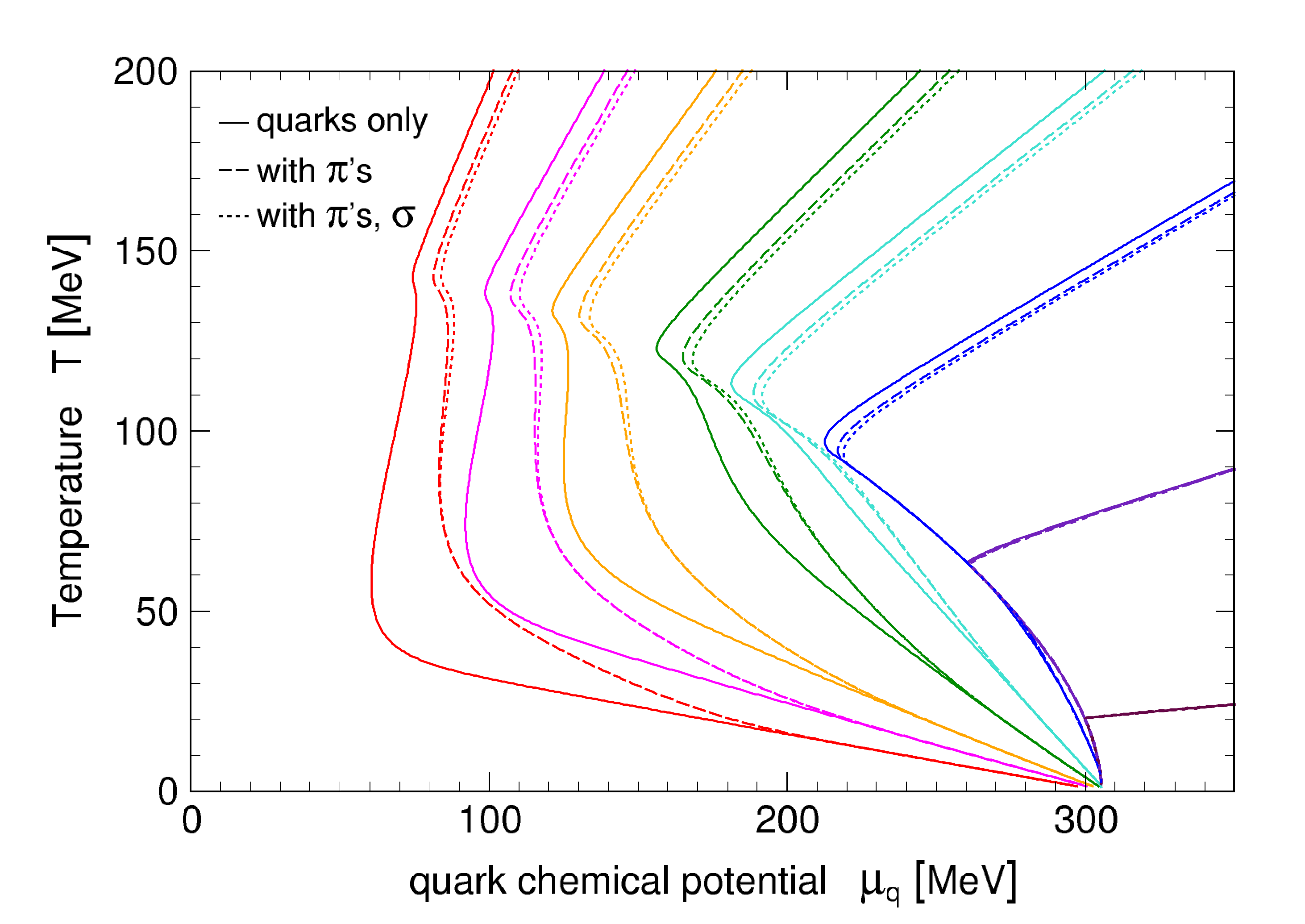}\hfill
	\begin{minipage}[b]{0.496\textwidth}
		\caption{Impact of the contribution of mesons to the isentropic trajectories. The solid lines are those in the left panel of \Fig{fig:isentropes_QM_NJL_Nf2} and with quarks only, the dashed lines are with an additional contribution of a pion gas to thermodynamics and the dotted line with a pion and $\sigma$-meson gas contribution.}
		\label{fig:isentropes_QM}
	\end{minipage}
\end{figure}

The considerable impact of the mesonic contribution to thermodynamics on isentropic trajectories is shown in \Fig{fig:isentropes_QM} in a QM model calculation, see also \Refs{Nakano:2009ps,Skokov:2010uh} for calculations using the Functional Renormalization Group.

\section{Impact of the number of quark flavours}

\begin{figure}
	\includegraphics[width=0.496\textwidth]{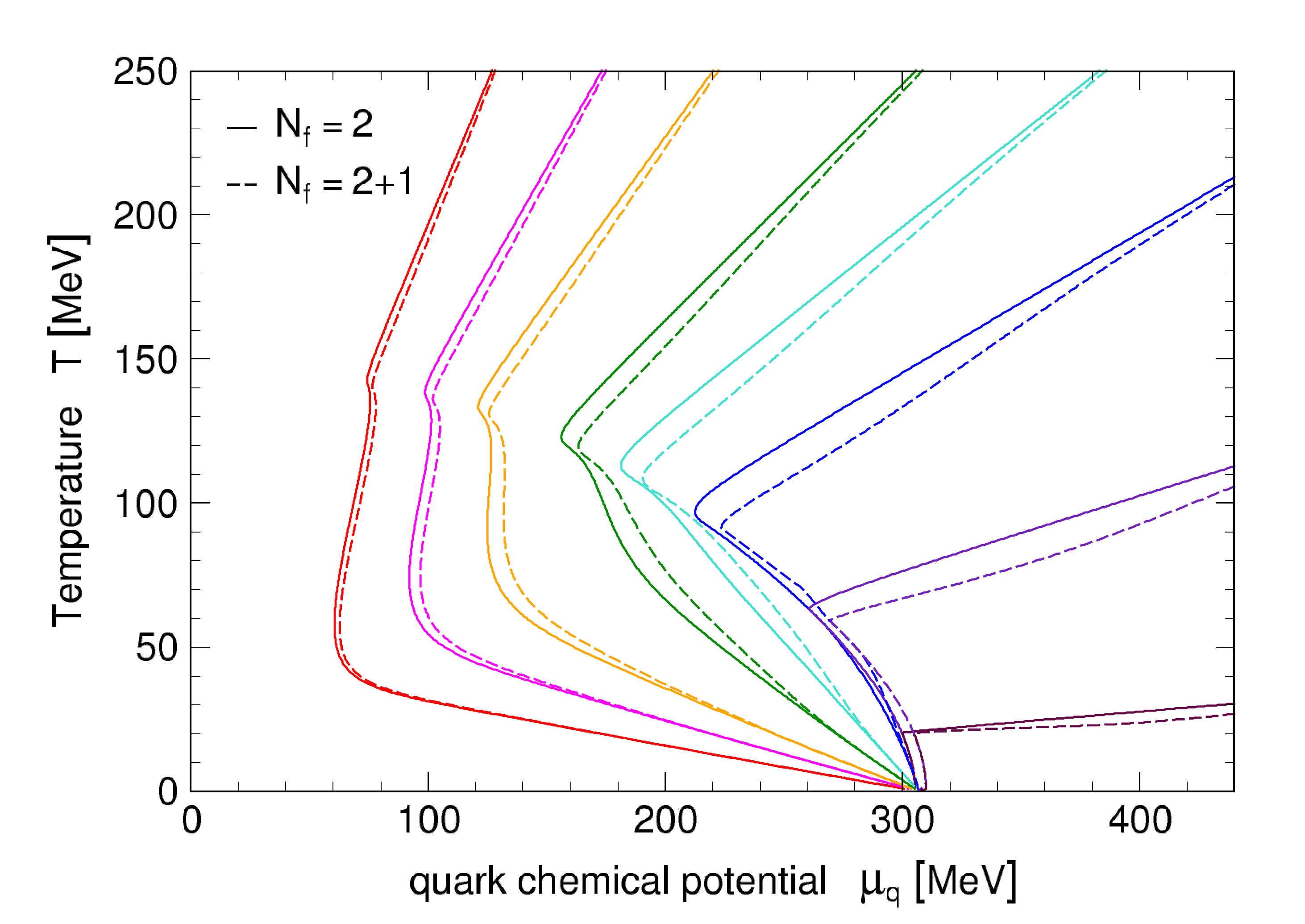}\hfill
	\begin{minipage}[b]{0.496\textwidth}
		\caption{Impact of the number of quark flavours on the isentropic lines. The solid lines are those in the left panel of \Fig{fig:isentropes_QM_NJL_Nf2} and the dashed ones the corresponding ones in a $N_\mrm{f}=2\!+\!1$\,--flavour calculation.}
		\label{fig:isentropes_Nf}
	\end{minipage}
\end{figure}

Isentropic lines in the PNJL model when strangeness is included were obtained in  \cite{Fukushima:2009dx,Costa:2010zw,Costa:2016vbb} for different scenarios.
Interestingly enough, the impact of including the strange quark in the calculation on the isentropic trajectories shown in \Fig{fig:isentropes_Nf} for the QM model is quite mild, less relevant than including the contribution of light mesons to thermodynamics, as shown in \Fig{fig:isentropes_QM}.

\section{Outlook}

A complete discussion of isentropic trajectories and of their potential to experimentally identify the presence of a critical end-point thanks to quantities like the speed of sound -- responsible for the collective acceleration of the fireball -- and quark-number susceptibilities -- connected to the fluctuations of conserved charges -- will be given in \Ref{Motta:2019}.
They will be also addressed within a systematic comparison between the (P)NJL and (P)QM frameworks in \Ref{Pereira:2019_1} and the impact of an improvement of the Polyakov-loop potential \cite{Baillot:2019} will be analysed.
The importance of lines of constant entropy to analyze the low-temperature/high-density phase within the Functional Renormalization Group approach will be discussed in \Ref{Pereira:2019_2}.
A discussion on finite size effects on isentropic trajectories can be found in \Ref{Palhares:2009tf}.

\ack
This work was supported by an INFN Post Doctoral Fellowship (competition INFN notice n.~18372/2016, RS) as well as by STSM Grants from the COST Action CA15213 ``Theory of hot matter and relativistic heavy-ion collisions'' (THOR) (RCP, PC, HH, MM, RS) and we thank for support and hospitality of CFisUC (project UID/FIS/04564/2016 - FCT Portugal).

\section*{References}
\bibliographystyle{iopart-num}
\bibliography{references}

\end{document}